\documentclass[
aps,
pra,
longbibliography,
superscriptaddress,
 amsmath,amssymb,
twocolumn,
]{revtex4-1}

\usepackage{graphicx}
\usepackage{dcolumn}
\usepackage{bm}
\usepackage{upgreek}
\usepackage{color}
\usepackage{hyperref}
\DeclareMathAlphabet{\mathpzc}{OT1}{pzc}{m}{it}
\usepackage{enumitem}   

\newcommand{\RR}{\right}
\newcommand{\LL}{\left}
\newcommand{\m}{\mathrm}
\newcommand{\dg}{\dagger}
\newcommand{\eref}[1]{Eq.~(\ref{#1})}
\newcommand{\fref}[1]{Fig.~\ref{#1}}

\usepackage[normalem]{ulem}

\setlist[enumerate]{itemsep=-1mm}

\begin{document}

\title{Optomechanics driven by noisy and narrowband fields }








\author{Louise Banniard}
\affiliation{Department of Applied Physics, Aalto University, FI-00076 Aalto, Finland}

\author{Cheng Wang}
\affiliation{Department of Applied Physics, Aalto University, FI-00076 Aalto, Finland}

\author{Davide Stirpe}
\affiliation{Department of Science and Industry Systems, University of South-Eastern Norway, PO Box 235, Kongsberg, Norway}

\author{Kjetil Børkje}
\affiliation{Department of Science and Industry Systems, University of South-Eastern Norway, PO Box 235, Kongsberg, Norway}

\author{Francesco Massel}
\affiliation{Department of Science and Industry Systems, University of South-Eastern Norway, PO Box 235, Kongsberg, Norway}

\author{Laure Mercier de L\'epinay}
\affiliation{Department of Applied Physics, Aalto University, FI-00076 Aalto, Finland}

\author{Mika A. Sillanp\"a\"a}
 \email{Mika.Sillanpaa@aalto.fi}
\affiliation{Department of Applied Physics, Aalto University, FI-00076 Aalto, Finland}

\date{\today}

\begin{abstract}

We report a study of a cavity optomechanical system driven by narrow-band electromagnetic fields, which are applied either in the form of uncorrelated noise, or as a more structured spectrum. The bandwidth of the driving spectra is smaller than the mechanical resonant frequency, and thus we can describe the resulting physics using concepts familiar from regular cavity optomechanics in the resolved-sideband limit. With a blue-detuned noise driving, the noise-induced interaction leads to anti-damping of the mechanical oscillator, and a self-oscillation threshold at an average noise power that is comparable to that of a coherent driving tone. This process can be seen as noise-induced dynamical amplification of mechanical motion.  However, when the noise bandwidth is reduced down to the order of the mechanical damping, we discover a large shift of the power threshold of self-oscillation. This is due to the oscillator adiabatically following the instantaneous noise profile. In addition to blue-detuned noise driving, we investigate narrow-band driving consisting of two coherent drive tones nearby in frequency. Also in these cases, we observe deviations from a naive optomechanical description relying only on the tones' frequencies and powers.


\end{abstract}

\maketitle

\section{Introduction}

Micro- and nanomechanical devices are central components in technology, and have become an indispensable tool for exploring fundamental physics, and in particular, quantum mechanics on a macroscopic scale. Mechanical oscillators coupled to high-frequency cavity modes (optical or microwave) have enabled the realization of various optomechanical interaction regimes under different driving conditions, exhibiting a plethora of intriguing physics and potential applications, including optomechanical cooling, amplification,  the manifestation of non-reciprocal effects, and the generation of non-classical quantum states.

In the vast majority of current cavity optomechanical systems, achieving strong optomechanical coupling between the optical and mechanical modes relies on coherent pumping of the system with an electromagnetic field. Since there are no sources producing purely coherent electromagnetic waves, noise in the driving field has to be considered in practice. In microwave cavity optomechanics, bandpass filtering is often required to purify the signal of interest.

Since this technical noise is generally spread on a broad band of frequencies, in noisy, non-filtered situations, its total integrated power can start to be comparable to the power of coherent tones employed to promote linear optomechanical interactions. Electromagnetic noise can therefore be expected to have a significant impact on optomechanical devices. To gain insight on the role of coherence in optomechanical devices, one can envisage the interesting situation where the electromagnetic driving of an optomechanical system is entirely replaced by electromagnetic noise with a certain spectrum, instead of a coherent driving tone. How would this influence the evolution of the mechanical system's dynamics? This question has been considered theoretically in the special case of a noisy driving field concentrated around the mechanical sideband red-detuned from the cavity \cite{Mari2012CoolHeat,Naseem2021coolheat}, and recently addressed experimentally in Ref.~\cite{2023_noisecool}. In the latter work, ground-state cooling of the oscillator with a noisy driving field was demonstrated.

A natural extension of the experimental study in Ref.~\cite{2023_noisecool} is to apply the noisy driving field around the blue mechanical sideband. If driven by a coherent tone, in this situation the optomechanical dynamical backaction causes familiar antidamping and eventually self-oscillation in the system \cite{Kippenberg2005LasingPRL,Marquardt2006MultiStab,Painter2010Laser,Lipson2012sync,Buks2014PRL,Painter2015oscill}. However, the lack of coherence in a noisy drive can be expected to result in randomly triggered incursions into the self-oscillation regime and therefore to interesting dynamics close to the instability threshold. 

In this work, we first of all implement such a blue-noise-driven experiment in a microwave cavity optomechanical system. We show that, in qualitative analogy with the case of coherent driving, driving an oscillator with noise around the blue sideband applies a heating optomechanical dynamical backaction on the oscillator and a mechanical instability occurs when the total power reaches a threshold value. However, we show that the instability threshold now depends on the noise bandwidth, a phenomenon that does not have an equivalent in the coherent case. To suggest a physical understanding of this phenomenon, we also inspect other types of spectral profiles of the driving field: two coherent tones around either sideband, essentially giving rise to a driving slowly modulated in time. This (otherwise coherent) situation resembles the incoherent driving case, although it involves only one typical drive modulation time.


\begin{figure*}[t]
    \centering
    \includegraphics[width=17cm]{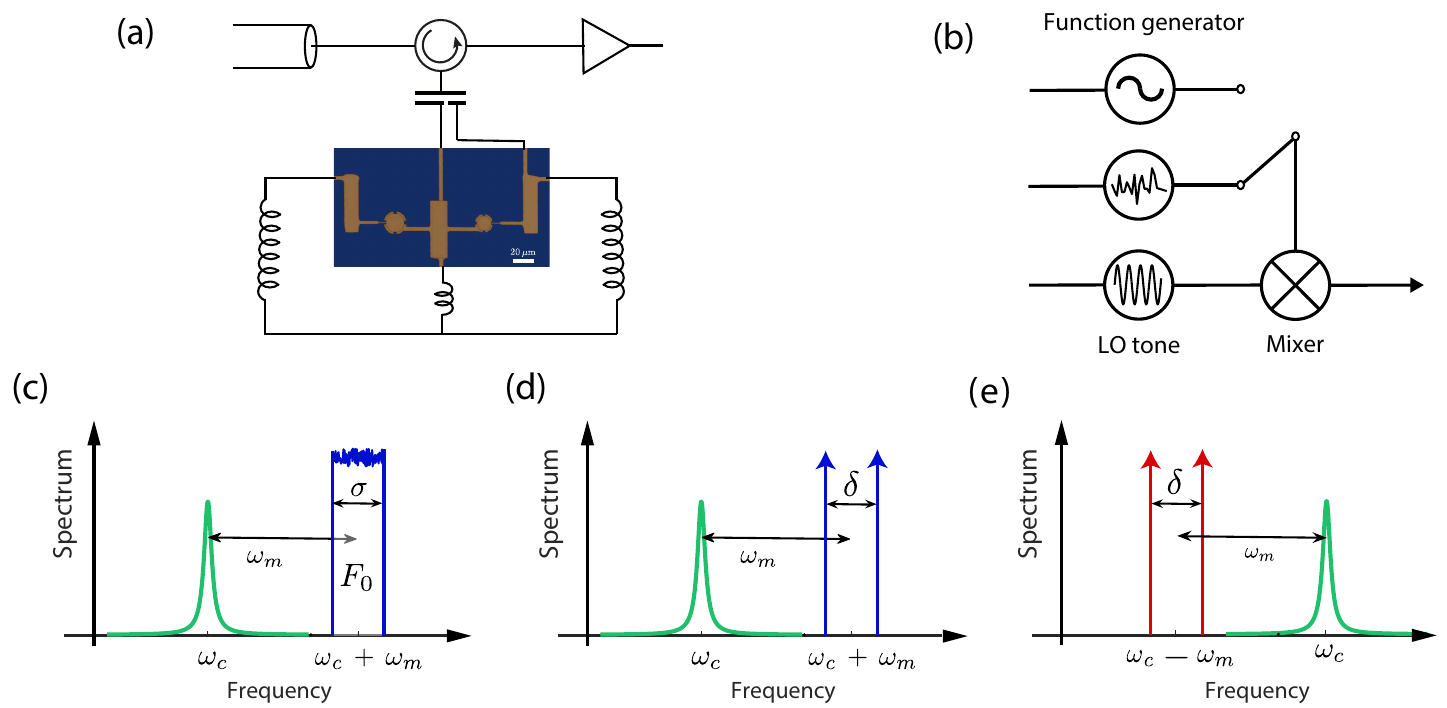} 
     \caption{\textit{Measurement scheme and device schematics.} (a) Simplified equivalent circuit of our microwave system and electromechanical device. The electromechanical circuit consists of two Al drumhead oscillators (out of which one is utilized), as shown in the center optical image,  both coupled to two microwave cavity modes formed by the combination of circuit inductors and capacitors. (b) The drive signal is created by mixing low frequency noise or a sine wave, generated with a function generator, with a local oscillator signal at frequency $\omega_c \pm \omega_m$ ($+$ blue-detuned, $-$ red-detuned). (c) Band-limited noise at the blue sideband, $\sigma$: noise width and $F_{0}$ noise flux. (d) Two tones around the blue sideband, $\delta$: frequency spacing. (e) Two tones around the red sideband.  
     } 
     \label{fig:figure1}
\end{figure*} 


\section{Models of narrowband driving}

\subsection{Optomechanical model}

The setup we consider is an optomechanical device, where a mechanical oscillator interacts parametrically with an electromagnetic resonator. 
For a detailed review of cavity optomechanics, the reader is referred to Refs.~\cite{OptoReview2014,BowenBook}. Here, we provide a non-exhaustive background which aids the reader to follow the rest of the paper. 
The frequency of the mechanical oscillator is denoted as $\omega_m$ and its damping rate as $\gamma$. The cavity frequency is denoted by $\omega_c$ and its damping rate by $\kappa$. The interaction is of radiation-pressure type, where the oscillator's position affects the cavity frequency with a coupling $g_0$:
\begin{align}
\label{eq:H0} 
H/\hbar = \omega_c a^\dg a + \omega_m b^\dg b + g_0 a^\dg a (b^\dg + b) \,.
\end{align}
Here, operators $a^\dg$ and $a$ are the harmonic oscillator ladder operators for the cavity, and similarly $b^\dg$ and $b$ are those of the mechanical oscillator.

In the canonical situation, the cavity is pumped strongly with a sinusoidal coherent tone of frequency $\omega_p$, inducing a cavity population of $n_0$ photons. The detuning of the pump tone from the cavity frequency is defined as $\Delta = \omega_p-\omega_c$. Under the strong pumping, it is possible to linearize the interaction term in \eref{eq:H0}, which becomes $ \hbar G (a^\dag +a) (b^\dag + b) $, where the effective coupling is $G = g_0 \sqrt{n_0}$. In the resolved-sideband regime, $\omega_m \gg \kappa$, the situation where the pump tone is applied close to the red-optomechanical sideband $\Delta=-\omega_m$, and the situation where it is applied close to the blue sideband $\Delta=+\omega_m$, promote different effective interactions, giving rise to different dynamical backactions. Assuming $|G| \ll \kappa$, the damping of the oscillator changes by the amount
\begin{align}
\label{eq:Gopt} 
\Gamma_\mathrm{opt} = \pm \frac{4 G^2}{\kappa} = \pm \frac{4 g_0^2 n_0}{\kappa} \,,
\end{align}
where the plus sign refers to red-sideband case, and the minus sign to the blue-sideband case. The total effective damping is thus $\Gamma_\m{eff} = \gamma + \Gamma_\mathrm{opt}$. The oscillator's phonon occupancy $n_m$ exhibits an associated cooling or heating,
\begin{equation}
\label{eq:nm}
n_m = \frac{\gamma \, n_m^T +\Gamma_\m{opt} \, n_{ba}}{\Gamma_\m{eff}}  \,,
\end{equation}
where $n_m^T \gg 1$ is the equilibrium thermal occupancy, and  $n_{ba}$ is a small correction due to finite sideband resolution. Based on \eref{eq:nm}, one also sees that the oscillator becomes unstable under the blue-sideband pumping as soon as a sufficient effective coupling is reached such that $\Gamma_\m{eff} \leq 0$.

\subsection{Cooling with noise}

The situation considered theoretically in Refs.~\cite{Mari2012CoolHeat,Naseem2021coolheat}, and very recently experimentally in Ref.~\cite{2023_noisecool} is that instead of coherent driving at the red sideband, the cavity is pumped at the red sideband with band-limited white noise. Under noise driving and with the bandwidth much larger than the resulting dynamical backaction damping, the dynamics is solved using the quantum-noise approach \cite{Marquardt2007,GirvinReview,Ojanen2014,OptoReview2014,BowenBook} (QNA). For cooling, the favored spectral profile of the injected noise is a box-function centered at the red sideband, with the width $\sigma$, flat-top value $N_0$, and the total flux $F_0 = \sigma N_0$. We also denote the average photon number induced by the driving in the cavity as $\bar{n}_0 \simeq \frac{\kappa F_0}{\omega_m^2}$ under the situations considered below. The added optical damping is 
\begin{equation}
\label{eq:goptnoise}
\gamma_{\m{opt}}^{\m{rsb}} = \frac{4 g_0^2 F_0}{\omega_m^2} \frac{\kappa}{\sigma} \tan^{-1} \left( \frac{\sigma}{\kappa}\right) \,, 
\end{equation}
and the oscillator is cooled according to \eref{eq:nm} with the replacement $\Gamma_{\m{opt}} \Rightarrow \gamma_{\m{opt}}^{\m{rsb}}$.
In the limit of a very small noise bandwidth, $\sigma \ll \kappa$, \eref{eq:goptnoise} (wrongly) predicts the familiar sideband-cooling result, \eref{eq:Gopt}. 

\subsection{Narrowband noisy field}

The regime of narrow band driving, specifically $\sigma \ll \gamma_{\m{opt}}^{\m{rsb}}$, exhibits interesting dynamics. A slowly varying time-dependent modulation of the driving field can adiabatically drive the oscillator's motion. The quantum noise model does not hold, but one can define a quasi-steady state of the oscillator in this narrow field driven regime.
This means that with a specific realization of the incoming amplitude waveform function denoted as $r(t)$, the average occupation is
\begin{align}
\label{eq:AvePhononOneRealization} 
 \bar{n}_m = \langle b^\dagger(t) b(t) \rangle_{r(t)} 
 = n_m^T \bigg\langle \frac{1 }{ 1+ q r^2(t)} \bigg\rangle_{r(t)} 
 \,.
\end{align}
In the case of incoming noise, the process $r(t) = r_n(t)$ is random, and follows a non-Gaussian distribution \cite{2023_noisecool}
\begin{align}
\label{eq:AmplitudeDist2} 
 P(r_n) =  \frac{2r_n}{F_0} e^{- \frac{r_n^2}{F_0}} \,,
\end{align}
and the parameter $q$ characterizes the transduction of the input field into the cavity. In the red-sideband centered pumping case, $q$ is given by
\begin{align}
\label{eq:qSBcool} 
  q_{\m{rsb}} & = \frac{4g_0^2}{\gamma \omega_m^2} \,.
\end{align}

Cooling is substantially inhibited in the narrow-band situation. The reason is that the average occupancy is dominated by periods of time when the instantaneous effective coupling is much smaller than the average, as the oscillator heats up during those periods of time in a manner is not compensated by periods of strong cooling.

Below, we consider three different driving schemes (see \fref{fig:figure1} c, d, e), focusing on the limit where the total spectral width is comparable to the mechanical linewidth including dynamical backaction. All the small-bandwidth cases can be treated based on \eref{eq:AvePhononOneRealization} with specific choices of $r(t)$ and the $q$ parameter. For convenience, we use the damping rate established by the average photon number as
\begin{equation}
\label{eq:goptbar}
\gamma_{\m{opt}} = \frac{4 g_0^2 \bar{n}_0}{\kappa} \,.
\end{equation}

\subsection{Noisy field at blue sideband}
\label{sec:bluenoise}

We now consider the driving noise spectrum shown in \fref{fig:figure1} (c), namely box-function noise profile centered at the blue sideband. Analogous to coherent tone at the same frequencies, we expect decreased damping of the mechanical oscillator. In the large bandwidth limit, analogous to red-sideband pumping, the negative contribution to mechanical damping is given by
\begin{align}
\label{eq:goptbsb} 
\gamma_{\m{opt}}^{\m{bsb}} = - \gamma_{\m{opt}}^{\m{rsb}} \,.
\end{align}
In the low bandwidth limit, 
\begin{align}
\label{eq:qSBcool} 
  q_{\m{bsb}} & = -\frac{4g_0^2}{\gamma \omega_m^2} \,,
\end{align}
and we may use \eref{eq:AvePhononOneRealization} to obtain the average phonon occupancy. However, now with negative added damping, the ensemble average in \eref{eq:AvePhononOneRealization} diverges for any nonzero values of optomechanical coupling. This is because a tail of the probability distribution even with a vanishing weight will trigger an infinite energy. In real life, of course, this is not the case, but the energy in the self-oscillation state is limited. In typical experimental situation, the self-oscillation amplitudes are limited by mechanical nonlinearities, and are in the range of $1 - 10$ nm \cite{OptoReview2014,2020_Grenoble}, corresponding to $n_m \sim 10^{10} - 10^{12}$. 

For a calculation, we thus introduce a capping value $n_{\m{max}}$ for self-oscillations. From the probability density function in \eref{eq:AmplitudeDist2}, one obtains the proportion of unstable events:
\begin{align}
\label{eq:Punstable} 
& P\LL(r_n > \frac{1}{\sqrt{q}} \RR)  = e^{-\frac{1}{q F_0}} = e^{-\frac{\gamma \omega_m^2}{4g_0^2 F_0}} =  e^{-\frac{\gamma}{\gamma_{\m{opt}}}} \,.
\end{align}
In the stable region, the population is typically not too much larger than $n_m^T$ and much smaller than in the self-oscillation regime. An estimate of the average population of blue-sideband noise driving in the narrow-bandwidth limit is thus given as 
\begin{align}
\label{eq:nunstable} 
& n_m \approx n_m^T + n_{\m{max}} e^{-\frac{\gamma}{\gamma_{\m{opt}}}}\,.
\end{align}

As discussed in connection with the experiments described below, we also simulate \eref{eq:AvePhononOneRealization} with random time series of the noise.

\subsection{Coherent tones around blue sideband}
\label{sec:bluecoh}

Let us now treat the case shown in \fref{fig:figure1} (d). Two equal coherent tones of amplitude $\alpha_{in}$ are applied symmetrically about the blue sideband frequency, with the spacing $\delta$. The input field is now
\begin{equation}
\begin{split}
& r(t) = 2 \alpha_{in} \sin \omega_{\m{bsb}} t \cos \delta t \,.
\label{eq:cohBSBinput}
\end{split}
\end{equation}
%

%
%
We expect a qualitatively similar dynamics whether the system is driven by a narrowband noisy field or a spectrum composed of coherent tones nearby in frequency. If the sideband processes due to the two tones do not overlap significantly, their dynamical backactions simply add up. The (anti)damping in the large frequency spacing limit is hence given by \eref{eq:goptbar}, with both $n_0$ and $\gamma_{\m{opt}}$ twice that of a single tone.

In the low frequency spacing limit, $\delta \ll \gamma_{\m{opt}}$, a common feature with narrow-band noisy driving is the slowly varying cavity field. In this regime,
\eref{eq:AvePhononOneRealization} gives
\begin{align}
\label{eq:nmBSBcohLowBW}
\bar{n}_m = \frac{\gamma n_m^T \kappa}{\sqrt{\gamma \kappa (\gamma \kappa - 8 g_0^2 \bar{n}_0 )}}
= \frac{\sqrt{\gamma} n_m^T }{ \sqrt{\gamma - 2 \gamma_{opt} }} \,,
\end{align}
where the average photon number is $\bar{n}_0  = \frac{\kappa \alpha_{in}^2}{\omega_m^2}$. The onset of instability is shifted towards lower average power as compared to a single tone because instantaneous values of $\gamma_{\m{opt}}$ can become larger for a given average power. This is qualitatively similar to noise driving at the blue sideband.

\subsection{Coherent tones around red sideband}
\label{sec:redcoh}

This situation is treated in the same fashion as in Section \ref{sec:bluecoh}, and the results are analogous. In the large-spacing limit, the damping is given by \eref{eq:goptbar} with the total photon number $\bar{n}_0$.

In the low-frequency spacing limit, we obtain as in \eref{eq:nmBSBcohLowBW}:
\begin{align}
\label{eq:nmRSBcohLowBW}
\bar{n}_m  = \frac{\gamma n_m^T \kappa}{\sqrt{\gamma \kappa (\gamma \kappa + 8 g_0^2 n_0)}}
  \approx n_m^T \sqrt{   \frac{ \gamma  }{ 2 \gamma_{opt} }} \,,
\end{align}
where the latter form holds in the strong damping limit. The cooling is thus suppressed as compared to a single tone, cf.~\eref{eq:nm}.

\section{Experiments}

\subsection{Device and signal generation}

\begin{figure}
    \centering
    \includegraphics[width=8.5cm]{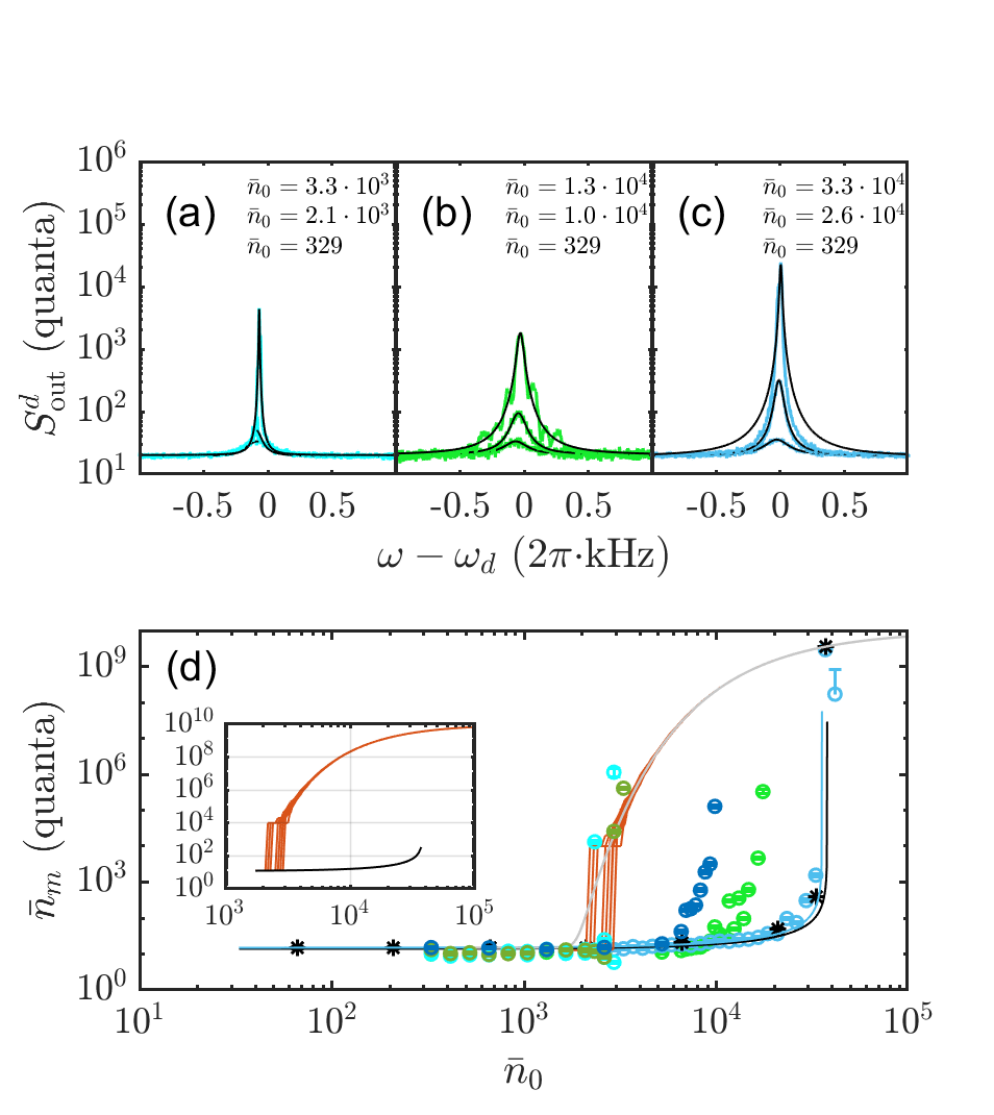}
    
     \caption{\textit{Optomechanical heating with blue-detuned noise.} Output spectra measured at the probe cavity with varied noise powers for three different noise bandwidths: (a) $\sigma = 2$ Hz, (b) $\sigma = 400$ Hz, and (c) $\sigma = 200$ kHz (Lorentzian fits in black). The photon numbers $\bar{n}_0$ are indicated in the caption from top to bottom. (d) The mechanical occupancy $n_m$ as function of photon number $\bar{n}_0$ for noise bandwidth $\sigma = 2$ Hz (cyan), $\sigma = 5$ Hz (dark green), $\sigma = 200$ Hz (dark blue), $\sigma = 400$ Hz (green), $\sigma = 200$ kHz (blue) and in black for the coherent driving tone. The solid blue line is the theory for 200kHz noise driving and the solid black line for the coherent tone case. In brown, we show the result of noise simulation iterated 20 times with a capping value $n_{\rm max}= 1\times10^{10}$, and in grey the corresponding theoretical prediction, \eref{eq:nunstable}. The inset illustrates the narrowband model over a large range.}
     
     
     \label{fig:figure2}
\end{figure} 

The device is a microwave circuit in aluminium whose two capacitors are formed by two drumhead oscillators (\fref{fig:figure1} (a)). In the experiments, only one of the two mechanical oscillators is utilized, with the fundamental mode frequency  ${\omega_m/2\pi \simeq 9.22\,\rm MHz}$ and the intrinsic damping rate $\gamma/2\pi \simeq 120$ Hz. The microwave circuit sustains two resonances at $\omega_{c}/2\pi \simeq 4.87\,\rm GHz$ and $\omega_{d}/2\pi \simeq 6.42\,\rm GHz$, both coupled to the mechanical oscillator in the resolved sideband regime ($\kappa < \omega_m$). The first cavity with decay rate of $\kappa/2\pi \simeq 1.06$ MHz used to drive the mechanical oscillator is called the "pump" cavity and the second cavity with decay rate of $\kappa_{d}/2\pi \simeq 0.84$ MHz used to characterize the mechanical oscillator is called the "probe" cavity. 


This dual-cavity circuit design greatly facilitates our experiment, especially in the case of noise driving in which the noise bandwidths used are sometimes larger than or of same order of the linewidth of the cavity. This requires another channel to access and probe the state of the oscillator, since the output spectrum from a noise-driven optomechanical cavity is not simply related to the mechanical spectrum in contrast to coherently driven cavity. 

\fref{fig:figure1} (b) illustrates the method that we use to create band-limited noises, or spectra consisting of two nearby coherent tones. Here, the method we adopt is straightforward and slightly different from that we used in our previous work where we inject noise at the red sideband. In case of blue-detuned noise, the baseband noise, spanning from zero frequency to $\sigma/2$, is first directly generated using a RF function generator. This noise is then up-converted by mixing with a LO tone at frequency $\omega_c+\omega_m$, creating a noise of bandwidth $\sigma$ located at blue sideband of the pump cavity, as shown in \fref{fig:figure1} (b).
The noise bandwidth $\sigma$ and the noise power quantified as averaged photon flux $F_0$ are two control parameters and can be varied to modify the noisy drive.

\begin{figure*}[t]
    \centering
     \includegraphics[width=15cm]{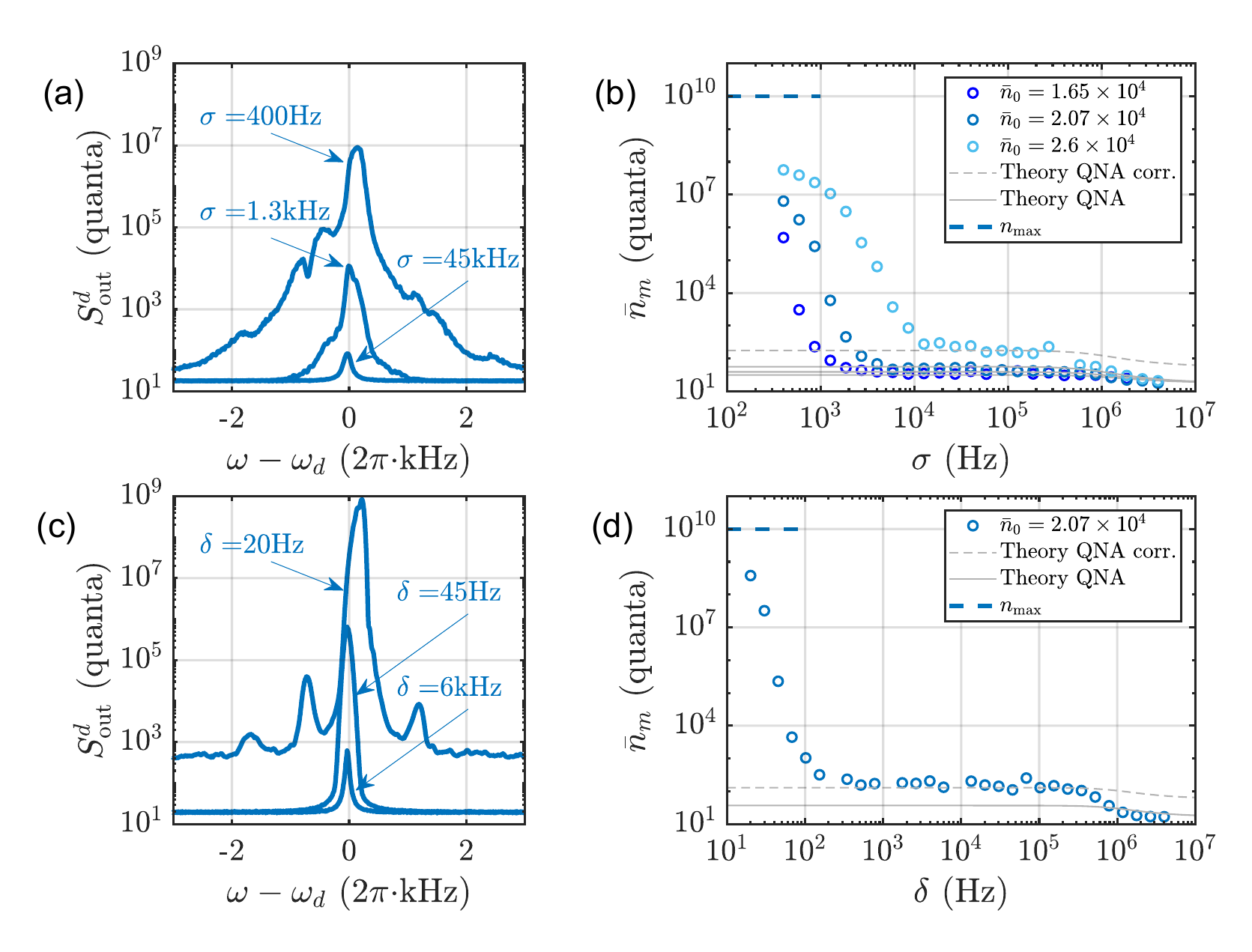}

     \caption{\textit{Self-oscillation induced by narrowband field driving.} 
     The panels (a) and (b) correspond to driving by blue-detuned noise [\fref{fig:figure1} (c)], while (c) and (d) correspond to driving with two tones centered at the blue sideband  [\fref{fig:figure1} (d)]. Typical spectra above and below the self-oscillation due to different noise drive bandwidths $\sigma$ are shown in panel (a), or frequency spacings 
     $\delta$ in panel (c). (b) The mechanical occupancy as a function of the noise bandwidth for three different noise fluxes. The solid grey lines represent the theory, with a correcting factor for the dashed grey line (see text). (d) The mechanical occupancy as a function of the two tones frequency spacing with the theory lines as an illustration of similarity between noise driving and two-tone driving.   }
     
     \label{fig:figure3}
\end{figure*} 
 
In the other measurement scheme, shown in \fref{fig:figure1} (d), two coherent tones with frequency spacing $\delta$ are evenly distributed on both sides of the blue optomechanical sideband frequency. The two-tone driving is created by mixing LO tone of frequency $\omega_c - \omega_m$ with a sine wave of frequency $\delta/2$  generated with same function generator. Finally, in \fref{fig:figure1} (e) we show the opposite situation, where two red-detuned tones are applied. In all cases, we also actively cancel the LO tone to make sure that the driving comes from the noise or the two tones alone. We also apply a weak probe tone to monitor and characterize the mechanical state by sending a tone at the red mechanical sideband of the probe cavity.

We utilized the same calibration method employed in our previous experiment on cooling by noise \cite{2023_noisecool}. The thermal occupancy of the mechanical oscillator resulting from the driving on the pump cavity can be well calibrated using a thermometry tone on the probe cavity. It is essential for this calibration that the mechanical mode can thermally equilibrate to the working temperature point at which the experiment is conducted. We ensure that our device can thermalize to the base temperature $\sim 10$ mK  of the cryostat by performing a thermal sweep measurement on the probe cavity. 


For the noise flux calibration, we cannot directly measure the photon flux $F_0$ that reaches the cavity due to the unknown attenuation factors in our measurement line. Instead, we perform a power sweep measurement using a coherent driving tone in place of the noise drive on the pump cavity. With the knowledge of single photon coupling strength ${g_{0}/2\pi \simeq 39\,\rm Hz}$ to the pump cavity, we can easily find the photon number $n_0$ populating the pump cavity at the given generator power. Next, we determine the power ratio between the narrow-band drive and the coherent tone settings at room temperature.
We thus can calibrate the averaged photon number $\bar{n}_0$ for the noise drive at the given generator output power.

\subsection{Driving under blue noise}

We first study the behavior of the system under blue-detuned noise driving.  
With a probe tone positioned at frequency $\omega_d-\omega_m$, the oscillator experiences backaction cooling, resulting in an effective linewidth $\gamma/2\pi= 220$ Hz 
and a mechanical phonon occupancy 
$n_m=12.1$. 
The injection of blue-detuned noise via the pump cavity leads to heating of the oscillator, accompanied by a narrowing of the mechanical linewidth and an increase in its power spectral density. This process can be seen as a noise-assisted dynamical backaction effect, similar to that observed in common cavity optomechanical systems driven by blue-detuned coherent tones, which leads to the amplification of the motion of the oscillator. 
As the drive power increases further above a certain threshold, the oscillator enters the regime of self-oscillation, onset of which is characterized by a rapid increase of population by orders of magnitude.


We chose to use several very different noise bandwidths and conducted power sweep measurements of the noise drive for each bandwidths.
By varying the power of the noisy drive at a fixed bandwidth $\sigma$, we monitor the mechanical state through probing with the red-detuned tone at the probe cavity. 
Modal spectra from the probe tone in the stable regime are presented in \fref{fig:figure2} (a), (b), and (c), each corresponding to a different noise bandwidth used. 
The mechanical occupancy $\bar{n}_m$ versus the calibrated averaged photon number $\bar{n}_0$ of the noise drive for the noise bandwith $\sigma=200 \rm kHz$ is well described by the heating predicted for blue-detuned coherent driving, as shown in \fref{fig:figure2} (d). However, when the bandwidth of the noise drive decreases and becomes comparable to the mechanical damping rate, we observed a clear deviation from the QNA result, Eqs.~(\ref{eq:goptnoise}, \ref{eq:goptbsb}). The power threshold, at which the mechanical self-oscillation occurs, shifts to a small value, and the shape of the measured spectra also starts to deviate from a Lorentzian. 
The measured spectral profile
can be modified as the bandwidth of noise spectrum changes, as observed between \fref{fig:figure2} (b) and (c).  Similar to the occupancy in this narrowband-fields-driven regime, the mechanical spectrum undergoes slow modulation, which imprints features to the time-averaged noise spectrum profile.

To analyse the noise-driven behavior in the narrowband regime, we fit the prediction in \eref{eq:nunstable} to the measured occupancy by using the self-oscillation energy $n_{\m{max}}$ as an adjustable parameter. The result displayed in \fref{fig:figure2} (d) shows a good agreement with the experiment, with a very reasonable value $n_{\m{max}} \simeq 10^{10}$.

It is worth looking at the approach to self-oscillations in a manner that corresponds more to the experimental situation, where the time series of the incoming random field $r_n(t)$ is of finite length. There are independent values roughly at the time intervals $\sigma^{-1}$ in \eref{eq:AvePhononOneRealization}. If the noise varies slowly, say at $\sigma/2\pi = 100$ Hz, and a typical spectrum scan takes on the order of a minutes, there are only $\sim 10^4$ points in the statistics. We thus further perform a simple simulation of \eref{eq:AvePhononOneRealization} with the knowledge of the noise statistics. In the simulation, we generate random samples to model the noise drive from the given probability distribution given by \eref{eq:AmplitudeDist2}. By performing an inverse transform sampling, the amplitude of noise drive can be modeled as $r_n= \sqrt{-\ln(1-u)/k}$, where $u$ is a sequence of uniformly distributed random numbers between 0 and 1. Through this method, we can obtain the mean phonon number $\bar{n}_m$ by averaging over the actual realization of the incoming noise. A capping value $n_{\m{max}}$ is utilized 
to represent the occupancy of self-oscillation by the oscillator, as in the analytical model [\eref{eq:nunstable}]. Then, we compute the average phonon occupancy across all instances of stable and unstable events for a particular noise sampling realization.

Based on the data shown in \fref{fig:figure2} (d), we observe that $\bar{n}_m$ versus $\bar{n}_0$ at $\sigma/2\pi$ in the few-Hz regime is well captured by simulation curves (brown) with the same capping value $n_{\rm max} =1 \times 10^{10}$ as used in the analytical model. There is jitter at the oscillation threshold due to different realizations of noise, because only the very largest and thus rarest noise values contribute to the onset. Equivalent variability is observed also in the experimental data.


Next, we keep the power of the drive field fixed and examine the bandwidth-dependent behavior with the  blue-detuned driving fields. 
%
The data displayed in \fref{fig:figure3} (a) and (b) depict the driving with blue-detuned noise. In (a), the mechanical spectrum at an intermediate bandwidth is well-behaved, but a self-oscillation occurs at the same noise flux at a small bandwidth as shown by the tall and broad spectrum. In \fref{fig:figure3} (b), we show the mechanical occupancy as a function of noise bandwidth. At large bandwidth ($\sigma > \kappa$), the system exhibits modestly suppressed heating consistent with the theoretical expectation. 

In the intermediate regime ($\gamma<\sigma<\kappa$) supposedly well described by the QNA, the heating is enhanced compared to large bandwidth limit and given by \eref{eq:goptbar}, without a clear bandwidth dependence. This regime exhibits a behavior comparable to our previous study on cooling with noise. The solid gray lines in \fref{fig:figure3} (b) describe the theory given by \eref{eq:goptbsb}, with a modest technical heating $n_m^T \simeq 32$ used as an adjustable parameter. A discrepancy relative to \eref{eq:goptbsb} occurs, however, for the largest noise power $\bar{n}_0 \simeq 2.6 \times 10^4$, which displays markedly enhanced heating. This can be due to the approaching of the onset of strong heating and instability in the QNA regime, where fluctuations in the device or physical environment may trigger additional excursions towards the unstable regime. This effect can be related to "spiking" of the mechanical oscillator \cite{Collin2019demag}. We model this as an enhanced technical heating up to $n_m^T \simeq 100$, shown by the dashed gray line in \fref{fig:figure3} (b).

As $\sigma$ becomes comparable to the oscillator's damping, we observe a strong heating of the oscillator, and when $\sigma$ is further reduced, the oscillator is driven into the regime of self-oscillation. The theoretical prediction from QNA, shown as solid lines in the figure, clearly fails to describe the behavior in the narrow-band regime. The occupancy is expected to approach the capping value $n_{\m{max}}$ marked in \fref{fig:figure3} (b) in the limit of vanishing bandwidth. This looks plausible, although based on this measurement, we cannot make a definite conclusion.

\begin{figure*}[t]
    \centering
    \includegraphics[width=14cm]{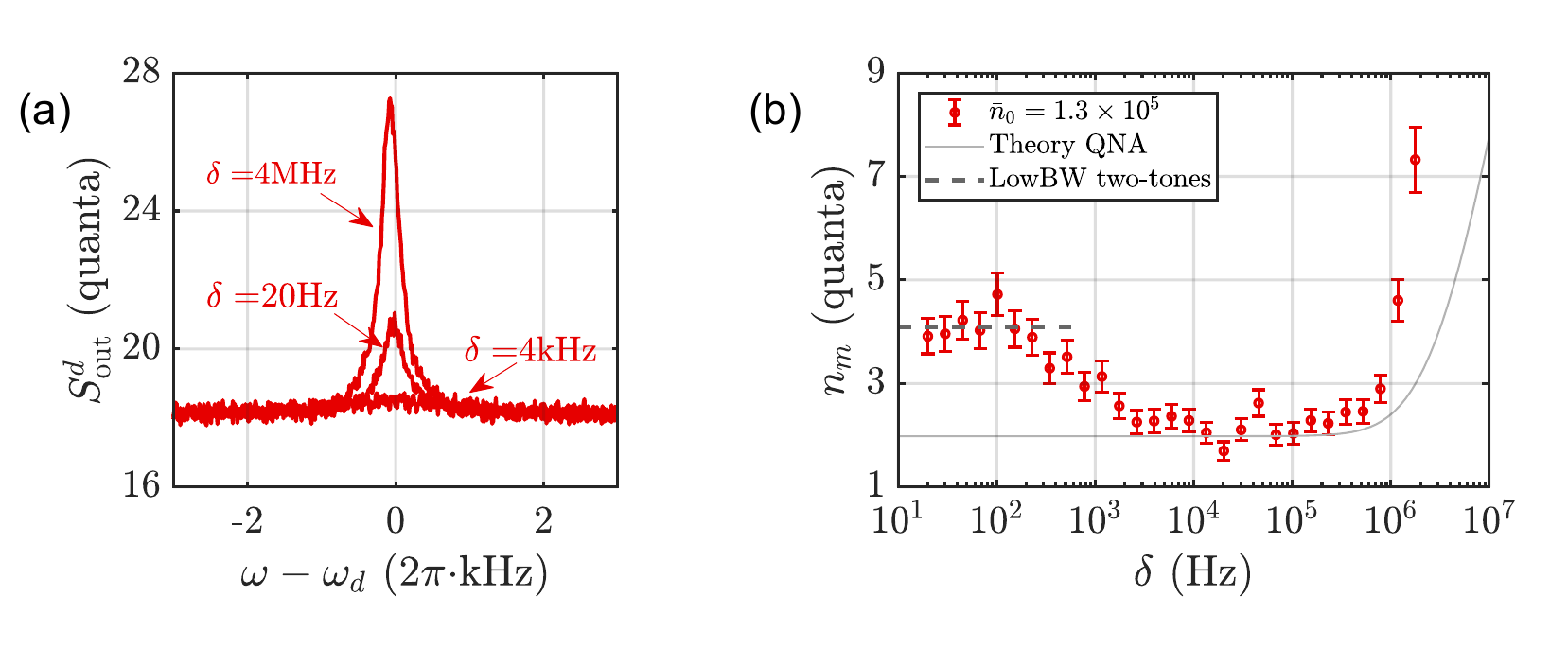} 
     \caption{\textit{Two-tone drive at the red sideband.} (a) Typical spectra at $\bar{n}_0 \simeq 1.3 \times 10^5$ for different two-tone spacings. (b) The mechanical occupancy as a function of the two-tone spacing. The grey dashed line represents the theoritical value for a two-tone drive with a small frequency spacing. } 
     \label{fig:figure5}
\end{figure*}

\subsection{Blue-detuned pair of coherent tones}

The second studied case of blue driving consists of two coherent tones situated symmetrically around the blue sideband (shown in \fref{fig:figure1} (d)). As with the noise-driven experiment, the probe cavity is used to characterize the oscillator, and the probe tone power and its associated damping were the same.

The result of a bandwidth sweep, where the power is constant but the frequency spacing between the tones is varied, is shown in \fref{fig:figure3} (c), (d). The result resembles the case of blue-noise driving: In the intermediate frequency spacing ($\gamma<\delta<\kappa$), we recover the standard dynamical backaction result, \eref{eq:Gopt}, while at large frequency spacing ($\delta > \kappa$), the heating is somewhat suppressed. And finally, small frequency spacings $\delta \lesssim \gamma$ are associated to strong heating and eventually self-oscillations. As compared to noise-driving, the oscillation threshold is shifted towards lower bandwidth due to a strict upper limit on the instantaneous flux. Similar to blue-noise driving, the heating at intermediate bandwidth is enhanced as compared to \eref{eq:Gopt}, again likely due to stronger technical heating.


\subsection{Red-detuned pair of coherent tones}

Finally, in the experiments we apply the pumping scheme in \fref{fig:figure1} (e), consisting of a pair of red-detuned tones. In analogy to cooling with noise \cite{2023_noisecool}, a narrow tone spacing is expected to inhibit cooling, according to \eref{eq:nmRSBcohLowBW}. The experimental data is displayed in \fref{fig:figure5}. In panel (b), the average occupancy extracted from the spectra in (a) is displayed as a function of tone spacing. We observe a reasonable agreement with the narrow-spacing prediction from \eref{eq:nmRSBcohLowBW}, shown by the dashed horizontal line. The solid line in (b) illustrates the large-bandwidth behavior, \eref{eq:goptnoise}, again showing similarity between red-detuned noisy field and a coherent field consisting of more than one frequency component.

\section{Conclusions}

In summary, we have investigated a cavity optomechanical system, which is driven by a narrowband electromagnetic spectrum. We focus on the situation where a driving on the optomechanical blue sideband, usually consisting of a single coherent tone leading to antidamping and eventual instability of the mechanical oscillator, is replaced by a band-limited white electromagnetic noise. We find that if the noise bandwidth is much larger than the mechanical loss rate but smaller than the cavity decay rate, the antidamping due to noise driving at a given cavity photon number equals that due to a pure sinusoidal drive. However, at narrow bandwidths, time-averaged antidamping due to noise driving substantially exceeds that of sinusoidal drive. We compare the results to another model of narrowband driving, that is, slowly amplitude-modulated coherent field, applied either at the blue or red sideband. Results qualitatively similar to noise driving are obtained in the latter cases, again with a large difference in slow-modulation limit in comparison to a single sinusoid. 

Our study can hold the potential to yield some insights into the behavior of complex systems, when the system is driven close to its instability. The well-known optomechanical self-oscillation, or parametric instability, occurring under coherent pumping at blue sideband has some analogy to lasing action, in the sense that an oscillation is triggered by pumping energy into the system. In our blue-noise driven experiment, the interpretation becomes stronger in the sense that the energy source is not a coherent pump, but resembles a heat source. Optomechanical self-oscillations can thus in principle be powered by temperature differences in the electromagnetic bath.


\begin{acknowledgments} We would like to thank Matthijs de Jong and Yulong Liu for useful discussions. We acknowledge the facilities and technical support of Otaniemi research infrastructure for Micro and Nanotechnologies (OtaNano). This work was supported by the Academy of Finland (contracts 307757, 312057), by the European Research Council (101019712), and by the Finnish Cultural Foundation. The work was performed as part of the Academy of Finland Centre of Excellence program (project 336810). We acknowledge funding from the European Union's Horizon 2020 research and innovation program under grant agreement 824109, the European Microkelvin Platform (EMP), and QuantERA II Programme (13352189). L. Mercier de Lépinay acknowledges funding from the Strategic Research Council at the Academy of Finland (Grant No. 338565). FM acknowledges financial support from the Research Council of Norway (Grant No. 333937) through participation in the QuantERA ERA-NET Cofund in Quantum Technologies.
\end{acknowledgments}


%

\end{document}